\documentclass{moriond}

\usepackage{amsmath,amssymb}

\def\be{\begin{equation}}
\def\ee{\end{equation}}
\def\bea{\begin{eqnarray}}
\def\eea{\end{eqnarray}}

\newcommand{\SO}{\mathrm{SO}}
\newcommand{\SU}{\mathrm{SU}}
\newcommand{\UH}{\mathrm{U}(1)_{\mathrm{H}}}
\newcommand{\U}{\mathrm{U}}

\newcommand{\Photo}{\includegraphics[height=35mm]{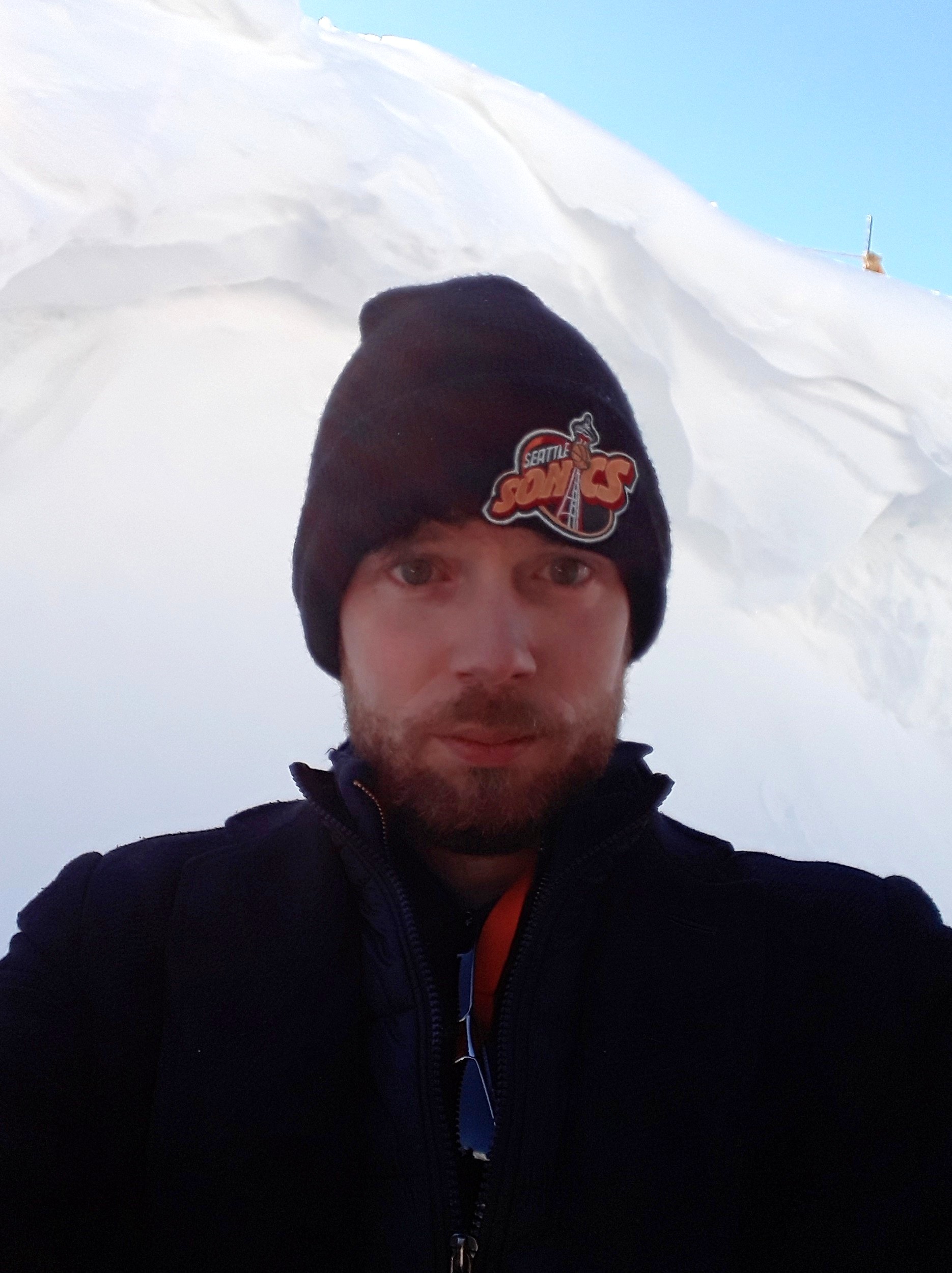}}

\begin{document}
\vspace*{4cm}
\title{AXIFLAVON-HIGGS UNIFICATION}

\author{TOMMI ALANNE, SIMONE BLASI, FLORIAN GOERTZ \footnote{Speaker}}

\address{Max-Planck-Institut f{\"u}r Kernphysik, Saupfercheckweg 1,\\ 
69117 Heidelberg, Germany \vspace{1.4mm}}

\maketitle\abstracts{
In this talk, a unified model of scalar particles that addresses the flavour hierarchies, solves the strong CP problem, delivers a dark matter candidate, and radiatively triggers electroweak symmetry breaking is discussed. The recently proposed axiflavon is embedded together with an (elementary) Goldstone Higgs-sector in a single multiplet (and thereby also a model of flavour and strong CP conservation for the latter is provided).
Bounds on the axion decay constant follow from requiring a SM-like Higgs potential at low energies and are confronted with constraints from flavour physics and astrophysics. In the minimal implementation, the axion decay constant is restricted to $f_a \approx (10^{11}-10^{12})$\,GeV, while adding right-handed neutrinos allows for a heavy-axion model at lower energies, down to $f_a \sim 10$\,TeV.}

\section{Introduction}

Although being very successful in describing nature 
in many aspects, the Standard Model (SM) of 
particle physics lacks explanations 
for several experimental facts, such as the 
significant abundance of Dark Matter, the large 
hierarchies in fermion masses and mixings, and the 
excellent conservation of CP symmetry in strong 
interactions, the latter being in tension with in principle 
sizable sources of CP violation in the QCD/SM Lagrangian. 
Beyond that, although the SM allows to parameterize
electroweak symmetry breaking (EWSB) via the Higgs 
mechanism, the origin of the Higgs potential remains
unexplained.

While separate solutions to all these problems are well known in the literature, here we entertain a unified model, simultaneously addressing all these problems at a time via a single scalar multiplet. To this end we show that it is possible to identify a radial component of this multiplet with a Froggatt-Nielsen-like flavon, whose vacuum expectation value (vev) now breaks an appropriate enhanced global symmetry $\mathcal{G}$ -- including a horizontal $\UH$ flavour symmetry -- down to $\mathcal{H} \not\supset \UH$. In turn, the corresponding (pseudo) Goldstone bosons of the $\mathcal{G}/\mathcal{H}$ coset can be identified with the axion and a (Goldstone-)Higgs doublet, as will be discussed in detail below. This leads to a unified description of symmetry breaking via a single fundamental source and unites the so-far separate scalar particles that solve the issues of massive EW gauge bosons (via {\it radiative} EWSB), of strong CP conservation, and of fermion mass\,+\,mixing hierarchies in one multiplet.

We start discussing our setup by reviewing briefly the
Froggatt--Nielsen (FN) mechanism~\cite{Froggatt:1978nt}.
This allows to address the fermion hierarchies by
charging the chiral SM fermions differently under a $\UH$
flavour symmetry. In turn, they are only allowed to 
interact with each other via a chain of new vector-like
fermions, $\xi_j$ (the FN messengers), connected via
insertions of a new complex scalar field $\Phi$, the 
flavon, carrying away the $\UH$-charge difference. 
After acquiring a vev, $\langle \Phi \rangle \neq 0$, 
the scalar spontaneously breaks the $\UH$ symmetry,
generating hierarchical Yukawa couplings, suppressed by
powers of its vev over the mass of the FN messengers, 
which are assumed to be somewhat heavier and thus 
integrated out in the IR theory, leaving us with 
effective interactions between the SM-fermion chiralities
(and the Higgs boson).  

In fact, it has been shown already~\cite{Calibbi:2016hwq,Ema:2016ops} that the angular component of $\Phi$ -- the axiflavon -- which plays no vital role in the original FN mechanism, can be identified with the QCD axion~\cite{Peccei:1977hh,Wilczek:1977pj,Weinberg:1977ma}, thereby addressing two more issues of the SM, namely the strong CP problem and the DM puzzle, in a unified scenario. This increases the predictivity of the axion solution, since the couplings of the latter are now dictated by the flavour structure, leading to interesting signals in flavour experiments~\cite{Calibbi:2016hwq}. Subsequently, it was shown that a further unification with the Higgs sector can be achieved~\cite{Alanne:2018fns}, as sketched above, increasing even more the predictivity of the scenario -- making this axion-like solution to the strong CP problem fully testable in the near future, while adding a dynamics to EWSB. Before discussing this framework in detail in the following, we finally add that, from a different perspective, it can also be seen as naturally including a model of flavour (and strong CP conservation) in the recently proposed elementary-Goldstone-Higgs scenario~\cite{Alanne:2014kea,Alanne:2015fqh}, furnishing a compelling renormalizable alternative to partial compositeness \cite{Kaplan:1991dc}.
  
\section{Explicit Model and Mass Hierarchies}

\label{sec:setup}

We embed the Higgs, the axion, and the
flavon in a single multiplet, $\Sigma$,
transforming under the enlarged symmetry group 
$\mathcal{G} \supset \UH \times G_\text{EW}$,
with $G_\text{EW} \equiv \SU (2)_{\mathrm{L}} \times \U (1)_Y$. Since the Higgs mass shall be much smaller than the
$\UH$ breaking scale, setting the flavon mass, we envisage both the axion and the Higgs boson components to
correspond to pseudo-Nambu--Goldstone bosons (pNGBs),
providing an example of axion-Higgs unification
\cite{Redi:2012ad} and allowing for dynamical EWSB via
the Coleman--Weinberg potential for the emerging
Goldstone Higgs. In fact, the vanishing of
the quartic Higgs coupling in the SM around $10^{10}$\,GeV, 
just about at the natural scale for axiflavon dark matter, 
might hint both to a Goldstone nature of the Higgs
and to a connection between the latter and the axion.

We thus formulate a linear sigma model for the field $\Sigma$ at the scale $f$, featuring Yukawa interactions with the FN messengers and the SM fields, with the (radial) flavon component developing a vev
breaking $ \mathcal{G} \xrightarrow{\text{\tiny $\langle \Sigma \rangle$}} \mathcal{H} \supset G_\text{EW}$, 
while the axion and the Higgs reside in the 
$\mathcal{G}/\mathcal{H}$ coset.
A particular simple choice for a viable breaking pattern is~\cite{Alanne:2018fns}
\begin{equation}\label{eq:so5pattern}
 \left[ \SO(5) \times \UH \right] \times \U(1)_X \rightarrow \SO(4)
 \times \U(1)_X\,,
\end{equation}
where the $\U(1)_X$ factor is introduced to reproduce 
the fermion hypercharges as $Y = T_3 + X$. 
The breaking above is obtained via a $\Sigma$ field living in the fundamental representation,
$\mathbf{5}$, of $\SO(5)$ and with $\UH$ flavour charge $H_\Sigma = 1$, featuring a potential
\begin{equation}\label{eq:potential}
 V(\Sigma, \Sigma^*) = 
 \lambda_1 \left(\Sigma^\dagger \Sigma \right)^2 - 
 \lambda_2 \, \Sigma^T \Sigma \, \Sigma^\dagger \Sigma^*  - \mu^2 \Sigma^\dagger \Sigma\,,
\end{equation}
which is bounded from below if $\lambda_1 > \lambda_2 > 0$.
The EW gauge group is embedded in $\SO(5)$ via the usual 
$\SU(2)_{\mathrm{L}} \times \SU(2)_{\mathrm{R}} \cong
\SO(4)$ generators~\cite{Alanne:2018fns} $T_{ij}^{a \, \mathrm{L,R}}$,
$a$\,=\,$1,..,3$, $i,j$\,=\,$1,..,4$. 
The $\SO(4)$-preserving minimum is then given by
$\langle \Sigma \rangle = (0,0,0,0,f/\sqrt{2})$, with
$ \mu^2 = (\lambda_1 - \lambda_2) f^2$. 

After the breaking of Eq.~\eqref{eq:so5pattern}, the scalar
sector can be parametrized as
\begin{equation}
\Sigma = e^{i (\sqrt2 h_{\hat a} \hat T^{\hat a} + a) /f} \begin{pmatrix} \widetilde{H} \\ (f + \sigma)/\sqrt 2 \end{pmatrix}\,,
\end{equation}
with the broken generators 
\begin{equation}
    \hat T_{ij}^{\hat a} = - \frac{i}{\sqrt 2} \left[ \delta_i^{\hat a}  \delta_j^5 - \delta_j^{\hat a}  \delta_i^5\right],
\end{equation}
where $\hat a$\,=\,$1,..,4$, and $i,j$\,=\,$1,..,5$. 
The physical states contained in the radial $\Sigma$-components are a heavy Higgs doublet, $\widetilde{H}$, with mass 
$m^2_{\tilde H} = 2 \lambda_2 f^2$, and a heavy flavon fluctuation
$\sigma$ around the vev $f$, with mass $m^2_\sigma = 2 (\lambda_1 - \lambda_2) f^2$, 
while the SM-like Higgs doublet, $h_{\hat a}$, and the axion, $a$, are instead pNGBs.

We assume that the explicit breaking of $\mathcal{G}$ originates from the SM sector only, namely from the QCD anomaly, EW gauging, and via the SM fermions not filling full $\mathcal{G}$ representations, while the FN messengers always enter as complete representations. The latter are denoted by $\xi_j$, where the subscript refers to the $\UH$ charge, $H_{\xi_j}=j$. They transform in the spinorial representation, $\mathbf{4}$, of $\SO(5)$ (which allows for a very minimal FN messenger sector~\cite{Alanne:2018fns}) and are non-chiral under $\SO(5) \times \UH$.
The SM fermions, $q_{\mathrm{L}}^i$, $u_{\mathrm{R}}^i$, and $d_{\mathrm{R}}^i$
(with $i=1,2,3$), come in incomplete SO(5) representations   in the spinorial
\begin{equation}\label{eq:spurion}
 \Psi^i_{q_{\mathrm{L}}} = \Delta_{\mathrm{L}}^T q^i_{\mathrm{L}}, 
 \quad \Psi^i_{u_{\mathrm{R}}} = \Delta_u^T u^i_{\mathrm{R}},
 \quad \Psi^i_{d_{\mathrm{R}}} = \Delta_d^T d_{\mathrm{R}}^i,                  
\end{equation}
where the spurions, $\Delta_{\mathrm{L},u,d}$, feature background values
\begin{equation}
    \Delta_{\mathrm{L}} = \left( \begin{array}{cccc}
                    1 & 0 & 0 & 0\\
                    0 & 1 & 0 & 0 \\
                   \end{array} \right), \quad
    \Delta_{u} = \left( 
                                          0, 0, 1, 0
                                        \right), \quad
    \Delta_{d} = \left( 
                                          0,0,0,1
                                         \right)\,,
\end{equation}
that parametrize the explicit SO(5) breaking (while $\UH$ remains exact at the Lagrangian level).
The rows of these $2\times 4$ matrices correspond to the
fundamental of the weak gauge group, while 
the columns label the components of $\mathbf{4}$ of SO(5).
The $\UH$ charge of each $\Psi^i_f$ is chosen such that
the correct pattern of masses and mixings is reproduced --
the larger the charge difference between the left- and 
right-handed components of a given SM fermion, the more suppressed is the resulting mass term, see below.

The corresponding Lagrangian includes renormalizable
operators made out of $\Psi_f^i$, $\xi_j$, and 
$\Sigma$ allowed by symmetries and reads
\begin{equation}
\begin{split}
 - \mathcal{L} \, & = \, \sum_j \left(a_j \bar{\xi}_{j+1} \, \Gamma^\alpha 
 \, \Sigma_\alpha \, \xi_j + \text{h.c.} \right)
 + m_j \, \bar{\xi}_j \, \xi_j,\\
 &+  \sum_{i,f=q_{\mathrm{L}}, u_{\mathrm{R}}, d_{\mathrm{R}}} 
 z_i^f \, \bar{\Psi}^i_f \, \Gamma^\alpha 
 \, \Sigma_\alpha \, \xi_{\bar\jmath} + \tilde z_i^f \, \bar{\xi}_{\bar\jmath + 2}
 \, \Gamma^\alpha \, \Sigma_\alpha \, \Psi^i_f 
 + x \, \bar{\Psi}^3_{q_L} \, \Gamma^\alpha 
 \, \Sigma_\alpha \, \Psi^3_{u_R} + \text{h.c.}\,,
 \end{split}
\end{equation}
where $\Gamma^a$ are the matrices defining the
spinorial representation
and the dimensionless coefficients $a_j,z_i^f,
\bar z_i^f,x$ are all assumed to be of ${\cal O}$(1). 
The first line contains the interactions of the FN
messengers with the $\Sigma$-field and their 
(vector-like) mass terms, while the second line 
consists of Yukawa couplings involving the SM fermions
and the FN messengers, with $\bar\jmath\equiv 
\bar\jmath(f,i) = H_{f^i}\!-\!1$ such that the terms are 
$\UH$ invariant, where the last term allows for an
unsuppressed top~mass.

It is easy to see how this Lagrangian leads to FN-like mass hierarchies. For two chiral SM fermions $\Psi_{q_{\mathrm{L}}}^i$ and $\Psi_{u_{\mathrm{R}}}^j$
with a $\UH$ charge difference of $\delta_{ij} \equiv H_{q_\mathrm{L}^i} - H_{q_\mathrm{R}^j}$, a chain of at least $|\delta_{ij}|-1$ FN messengers $\xi_i$ together with $|\delta_{ij}|$ insertions of the $\Sigma$ field is required to couple them. After integrating out the $\xi_i$ at the tree level (and suppressing ${\cal O}(1)$ factors), one finds the corresponding effective leading order Lagrangian 
\begin{equation}\label{eq:Leff2}
 - \mathcal{L_{\text{eff}}} \sim \frac{1}{m^{|\delta_{ij}|-1}}
 \bar{\Psi}_{u_{\mathrm{R}}}^j
 (\Gamma^\alpha \Sigma_\alpha)^{|\delta_{ij}|}
 \, \Psi_{q_\mathrm{L}}^i + \text{h.c.} \,,
\end{equation}
which is non-vanishing for odd $|\delta_{ij}|$,
as assumed in the following, while for even 
$|\delta_{ij}|$ it is zero due to $\Gamma$-matrix
properties.

Below the symmetry-breaking scale and after integrating 
out the flavon and the second Higgs doublet, 
the $\Sigma$-field can be written 
via the Goldstone parametrization in the unitary gauge as
\begin{equation}\label{eq:sigmagoldstone}
 \Sigma = \frac{f}{\sqrt{2}} \, e^{i a/f} \left(
                          0, 0, \text{sin} \, h/f
                          , 0, \text{cos}\, h/f
                         \right)^T,
\end{equation}
where $h$ represents the Higgs field, and $a$ is the axion. With this, we can single out the leading contribution to the mass matrix
\begin{equation}\label{eq:Leffmass}
 - \mathcal{L_{\text{eff}}} \supset m_{ij} \bar{q}_{\rm L}^i u_{\rm R}^j  
 + \text{h.c.} \,, \quad
 m_{ij} \sim \frac{v}{\sqrt{2}} \left(\frac{f^2}{2 m^2}\right)^{\frac{|\delta_{ij}| - 1}{2}}\, ,
\end{equation}
where $v \equiv f \, \sin (\langle h\rangle/f)$ is the EW scale. This exhibits a typical FN suppression in
$\epsilon \equiv \left(f^2 / 2 m^2\right)$, which 
we identify with the Cabibbo angle, $\epsilon = \text{sin}\,\theta_{\mathrm{C}} \simeq 0.23$,
to reproduce the flavour hierarchies.

\section{Higgs Potential and Constraints on the Axion Decay Constant}
\label{sec:Pot}

To calculate the Higgs potential, we consider the top sector, with a charge assignment $H_{q_{\rm L}^3} = 1$, $H_{u_{\rm R}^3} = 2$, compatible with the top mass, and two (mass-degenerate) FN messengers, $\xi_{0,1}$ coupled to them. This allows for a minimal non-trivial chain of messengers and captures the leading effect~\cite{Alanne:2018fns}. The potential is then computed by matching at one loop the SM Coleman--Weinberg potential, renormalized at the mass scale of the FN messengers $m_0=m_1 \equiv m$, with the one in the axiflavon-Higgs scenario, induced by explicit SO(5) breaking interactions involving the fields mentioned above, see Ref.~\cite{Alanne:2018fns} for details. Requiring the quadratic term to reside at the electroweak scale $v\,(\ll\!f)$ allows us to make a \emph{prediction} for the quartic coupling at the scale $m$ in terms of the top Yukawa, which turns out to be very small and negative. It reads~\cite{Alanne:2018fns}
\begin{equation}\label{eq:lambda}
 \lambda(m) = - \frac{N_c}{ 2 \pi^2} \frac{f^2}{ 2 m^2} 
 y_t^6(m) (1+\delta)^6\, ,
\end{equation}
where following the assumptions of the FN mechanism, all dimensionless coupling parameters have been chosen to be of the order of the (${\cal O}(1)$) top Yukawa, which is explicitly realized by replacing them with an average Yukawa coupling $y_t(m)(1+\delta)$, including a spread of $|\delta|<1$.

\begin{figure}
\centering
 \includegraphics[scale=0.55]{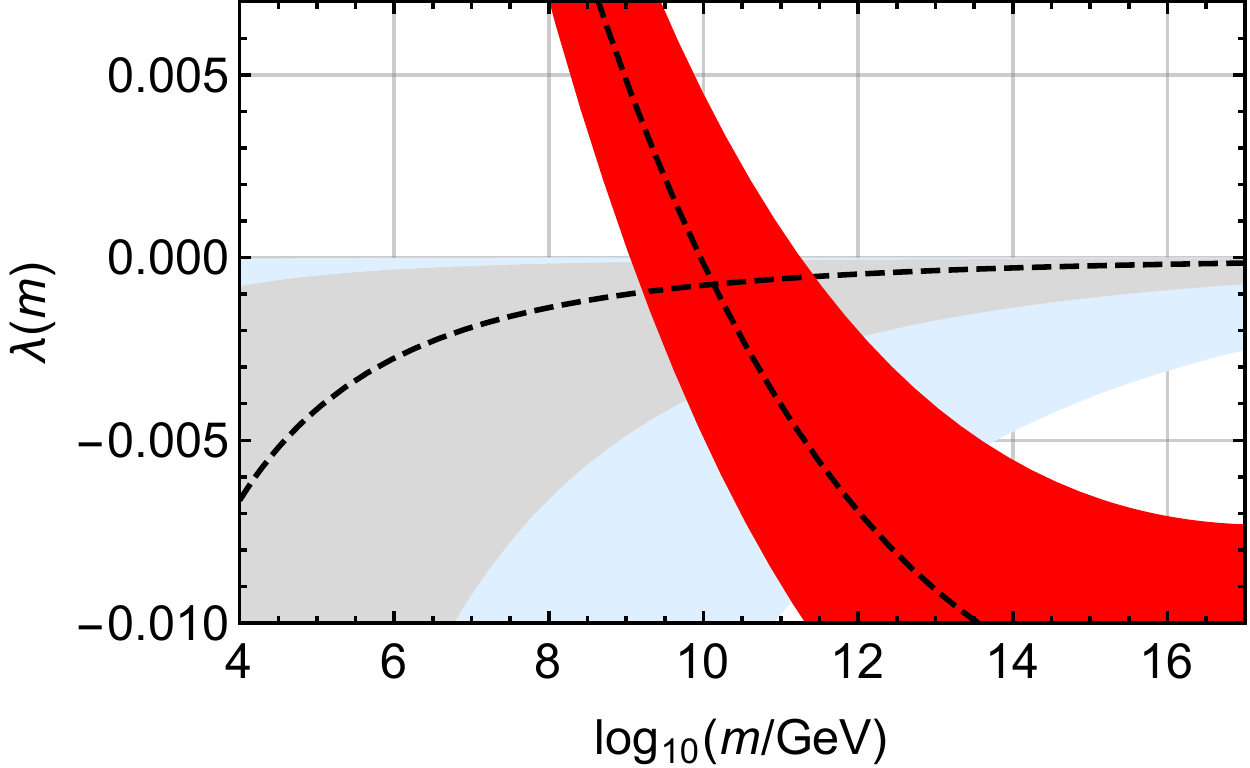}\qquad
 \includegraphics[scale=0.525]{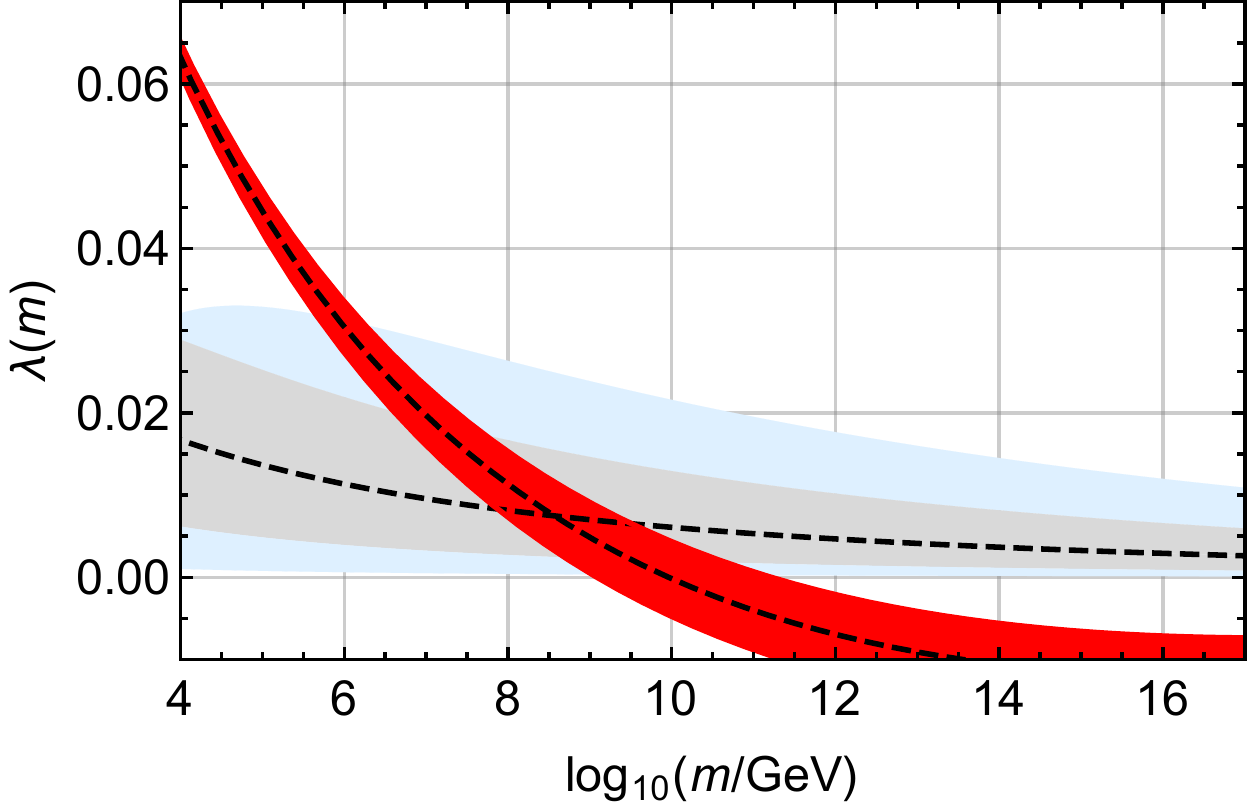}
 \caption{Left: Matching of the Higgs quartic coupling at the scale $m$ in the SM (red band) with the prediction of Eq.~(\ref{eq:lambda}), considering a Yukawa coupling
 spread of $\delta = \pm 0.6$ (light blue band),
$\delta = \pm 0.3$ (light gray band), and
$\delta = 0$ (dashed black line). The intersection corresponds to the allowed range for $m$. Right: Same plot for the low scale model, Eq.(\ref{eq:lambdan2}). See text for details.}
 \label{plots}
\end{figure}

Since below the threshold of the FN messengers $\lambda(m)$ is fully predicted via SM running, Eq.~\eqref{eq:lambda} can be used to determine the scale $m$ at which a successful matching is achieved.
In the left panel of Fig.~\ref{plots} we show the SM running of $\lambda(m)$ (including uncertainties) via the red band and the RHS of Eq.~\eqref{eq:lambda}
for $\delta = \pm 0.6$ (light blue band),
$\delta = \pm 0.3$ (light gray band), and
$\delta = 0$ (dashed black line). The matching is possible only for negative values of $\lambda(m)$, which selects $10^9 \, {\rm GeV} \lesssim m \lesssim 10^{14} \, {\rm GeV}$. This translates, via $f^2/ 2 m^2 \simeq 0.23$ and $f_a = f/N$, with
\begin{equation}
\label{eq:DWN}
 N = \sum_i 2 H_{\Psi_{q_L}^i} - H_{\Psi_{u_R}^i} - H_{\Psi_{d_R}^i} \approx 50,
\end{equation}
to a axion decay constant of
\begin{equation}\label{eq:bound1fa}
 10^7 \, \text{GeV} \lesssim \, f_a \, \lesssim 
 10^{12} \, \text{GeV}.
\end{equation}

\begin{figure}
\centering
 \includegraphics[scale=0.42]{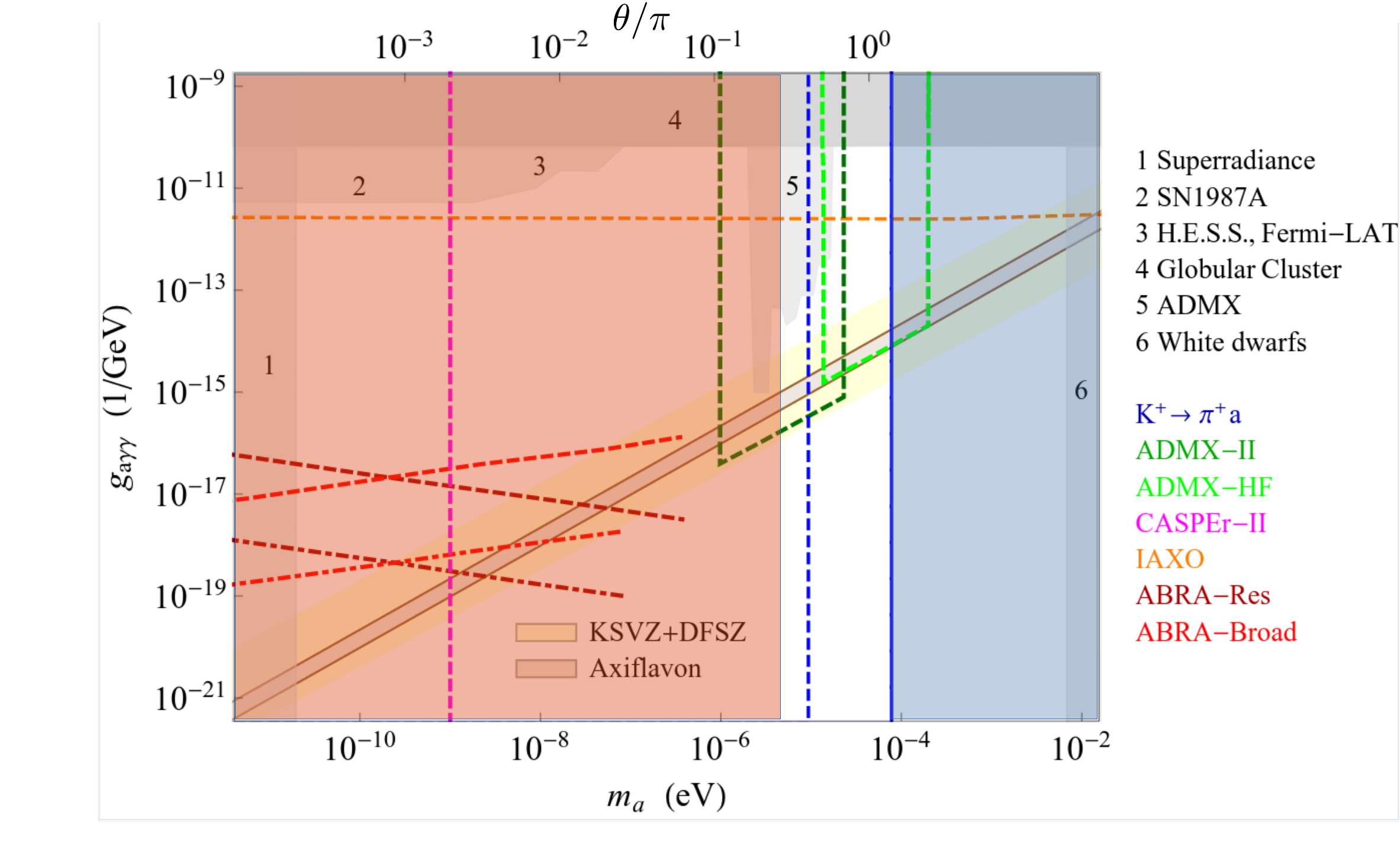}
 \vspace{-5mm} 
  \caption{\label{plot2} Constraints on the axion parameter space, i.e.\ axion mass vs.\ photon coupling, in the scenario considered. The axiflavon prediction is depicted by the thin brown band. The exclusions from various axion experiments, summarized in the legend, are given as grey numbered regions. The dashed colored lines show the projected reach of future experiments. On the other hand, the blue shaded region corresponds to the estimated bound from current flavour experiments, while the dashed blue line gives the expected reach of NA62. Finally, the red shaded area depicts the region that is excluded since a consistent matching of the Higgs potential is impossible (for the minimal model), which furnishes a new constraint due to axiflavon-Higgs unification.}
\end{figure}

This can be confronted with constraints due to the flavour-violating couplings of our axion~\footnote{Note that, for fixed $f_a$, the axion couplings to fermions and to photons remain basically unchanged compared to the original axiflavon model~\cite{Calibbi:2016hwq} (becoming exactly equal in the limit of large $\UH$ charges).
For example, due to the different $\UH$ charge assignment~\cite{Alanne:2018fns} ($|\delta_{ij}| \to 2 |\delta_{ij}| +1$), the color anomaly, Eq.~(\ref{eq:DWN}),
changes by $N \to 2N + 2 \approx 2 N$.
This translates (for constant $f_a$) into the same change in $f$, while also the electromagnetic anomaly changes by approximately a factor of two, canceling in the coupling to photons.}. Limits from searches for $K^+ \to \pi^+ a$ lead to $f_a \gtrsim 7.5 \times 10^{10}$\, GeV at $90\%$\,C.L. \cite{Calibbi:2016hwq}, leaving a relatively thin stripe of
\begin{equation}
\label{eq:cons}
 f_a \approx (10^{11}-10^{12}) \, \text{GeV}\,,
\end{equation} 
which will almost entirely be tested by the NA62 experiment, that just started operation~\cite{Dobrich:2018ezn}. The combined exclusions from requiring a consistent matching of the Higgs potential and satisfying flavour bounds are visualized in Fig.~\ref{plot2} as red and blue shaded regions, respectively, which shows the axion parameter space~\cite{Calibbi:2016hwq},
where $g_{a\gamma\gamma}$ is the axion-photon coupling and $m_a = 5.7\, \mu{\rm eV}\,(10^{12}\,{\rm GeV}/f_a)$~\cite{diCortona:2015ldu}.

We finally discuss the impact of including
right-handed (RH) neutrinos $N_{\mathrm{R}}$, which enter, together with the left-handed lepton doublet $l_{\mathrm{L}}$ as $\SO(5)$ spurions (see Eq.~\eqref{eq:spurion}) 
\begin{equation}
\Psi_{\mathrm{L}} = \Delta^T_{\mathrm{L}} l_{\mathrm{L}}, 
    \quad \Psi_N = \Delta_u^T N_{\mathrm{R}}.
\end{equation}
We assign flavour charge to $\Psi_N$ such that
the term
\begin{equation}
    \label{eq:majorana}
	- \mathcal{L}_N = \frac{1}{\sqrt{2}} y_N \bar{\Psi}_N \Sigma^\prime {\cal C} \bar{\Psi}_N^T + 
	 \text{h.c.}
	 =  -\frac{1}{2} y_N f\, \cos(h/f) \bar{N}_{\mathrm{R}} {\cal C}
	 \bar{N}_{\mathrm{R}}^T e^{\imath a/f} + \text{h.c.}
\end{equation}
is allowed, leading to Majorana and Dirac masses (the latter via a FN-chain)
\begin{equation}
    m_{N_{\mathrm{R}}}^2(h) = y_N^2 f^2\,\cos^2(h/f)\,, \quad
 m_{\mathrm{D}} \sim m_t \epsilon^{\frac{|\delta_\nu|-1}{2}},
\end{equation}
where $\delta_\nu = H_{l_L} - H_{N_{\mathrm{R}}}$.
The light neutrino mass is then given by
\begin{equation}
 m_\nu \sim m_t \epsilon^{|\delta_\nu|-1} \frac{m_t}{m_{N_{\mathrm{R}}}},
\end{equation}
which shows a double suppression,
originating from the type-I seesaw and from the FN mechanism.
Including the impact of three almost degenerate 
RH neutrinos, described by Eq.~\eqref{eq:majorana} with a typical coupling parametrized as $ y_N = (1 + \delta) y_t$, on the Higgs potential we arrive now (similar as before) at a positive
\begin{equation}\label{eq:lambdan2}
 \lambda(m) =  \frac{3}{8 \pi^2} 
 \log\left(\frac{1}{2 y_t^2(m) (1+\delta)^2 \epsilon}\right)
 (1 + \delta)^4 y_t^4(m).
\end{equation}

In the right panel of Fig.~\ref{plots} we confront the SM running of $\lambda(m)$ in red and the RHS of Eq.~\eqref{eq:lambdan2} for $\delta = \pm 0.6$ (light blue band), $\delta = \pm 0.3$ (light gray band) and
$\delta = 0$ (dashed black line).
The matching is now possible for smaller values of $m$ with respect to the case without RH neutrinos, leading to
\begin{equation}
\label{eq:cons2}
 6 \, \text{TeV} \lesssim f_a \lesssim 2 \times 10^{6} \, \text{TeV}\,.
\end{equation}
Although such low values of $f_a$ are excluded
for the usual QCD axion, by disentangling the axion mass and decay constant, low-$f_a$ models can 
become viable~\cite{Gherghetta:2016fhp,Gaillard:2018xgk}. 
Supernova cooling and flavour constraints can then be avoided by pushing the axion mass to the GeV or TeV scale. While this axion cannot be a dark matter candidate anymore, 
it still solves the strong CP problem and the RH neutrinos can add a link to the
matter-antimatter asymmetry via leptogenesis~\cite{Alanne:2017sip}.

\section{Conclusions}
\label{sec:con}
\vspace{-2mm}

A unified model that addresses the flavour puzzle, solves the strong CP problem, provides a dark matter candidate, and radiatively triggers electroweak symmetry breaking has been presented, employing
a single scalar multiplet which also contains the Higgs boson. The model is highly predictive, leading to the constraints on the axion decay constant presented in Eqs.~(\ref{eq:cons}) and (\ref{eq:cons2}) for the minimal incarnation and the model including right handed neutrinos, respectively, with the relevant (current and projected) bounds for the former summarized in Fig.~\ref{plot2}.

\section*{Acknowledgments}
\vspace{-3mm}

FG would like to thank the organizers of the
Rencontres de Moriond 2019 EW for the invitation and the wonderful atmosphere during the conference.


\section*{References}
\vspace{-3.5mm}

\begin{thebibliography}{99}

\bibitem{Froggatt:1978nt}
  C.~D.~Froggatt and H.~B.~Nielsen,
  Nucl.\ Phys.\ B {\bf 147} (1979) 277.

\bibitem{Calibbi:2016hwq}
  L.~Calibbi, F.~Goertz, D.~Redigolo, R.~Ziegler and J.~Zupan,
  Phys.\ Rev.\ D {\bf 95} (2017) no.9,  095009

\bibitem{Ema:2016ops}
  Y.~Ema, K.~Hamaguchi, T.~Moroi and K.~Nakayama,
  JHEP {\bf 1701} (2017) 096

\bibitem{Peccei:1977hh}
  R.~D.~Peccei and H.~R.~Quinn,
  Phys.\ Rev.\ Lett.\  {\bf 38} (1977) 1440.

\bibitem{Wilczek:1977pj}
  F.~Wilczek,
  Phys.\ Rev.\ Lett.\  {\bf 40} (1978) 279.

\bibitem{Weinberg:1977ma}
  S.~Weinberg,
  Phys.\ Rev.\ Lett.\  {\bf 40} (1978) 223.
  
\bibitem{Alanne:2018fns}
  T.~Alanne, S.~Blasi and F.~Goertz,
  Phys.\ Rev.\ D {\bf 99} (2019) no.1,  015028

\bibitem{Alanne:2014kea}
  T.~Alanne, H.~Gertov, F.~Sannino and K.~Tuominen,
  Phys.\ Rev.\ D {\bf 91} (2015) no.9,  095021

\bibitem{Alanne:2015fqh}
  T.~Alanne, A.~Meroni, F.~Sannino and K.~Tuominen,
  Phys.\ Rev.\ D {\bf 93} (2016) no.9,  091701

\bibitem{Kaplan:1991dc}
  D.~B.~Kaplan,
  Nucl.\ Phys.\ B {\bf 365} (1991) 259.
  

\bibitem{Redi:2012ad}
  M.~Redi and A.~Strumia,
  JHEP {\bf 1211} (2012) 103

  
  
\bibitem{Dobrich:2018ezn}
  B.~D{\"o}brich [NA62 Collaboration],
  Frascati Phys.\ Ser.\  {\bf 66} (2018) 312

\bibitem{diCortona:2015ldu}
  G.~Grilli di Cortona, E.~Hardy, J.~Pardo Vega and G.~Villadoro,
  JHEP {\bf 1601} (2016) 034

\bibitem{Gherghetta:2016fhp}
  T.~Gherghetta, N.~Nagata and M.~Shifman,
  Phys.\ Rev.\ D {\bf 93} (2016) no.11,  115010
  
\bibitem{Gaillard:2018xgk}
  M.~K.~Gaillard, M.~B.~Gavela, R.~Houtz, P.~Quilez and R.~Del Rey,
  Eur.\ Phys.\ J.\ C {\bf 78} (2018) no.11,  972

\bibitem{Alanne:2017sip}
  T.~Alanne, A.~Meroni and K.~Tuominen,
  Phys.\ Rev.\ D {\bf 96} (2017) no.9,  095015
  
\end{thebibliography}
\end{document}